# Feedback Consolidation Algorithms for ABR Point-to-Multipoint Connections in ATM Networks*

Sonia Fahmy, Raj Jain, Rohit Goyal, Bobby Vandalore, and Shivkumar Kalyanaraman
The Ohio State University
E-mail: {fahmy,jain}@cis.ohio-state.edu
Sastri Kota
Lockheed Martin Telecommunications
Pradeep Samudra
Samsung Telecom America, Inc.

**Abstract:** ABR traffic management for point-to-multipoint connections controls the source rate to the minimum rate supported by all the branches of the multicast tree. A number of algorithms have been developed for extending ABR congestion avoidance algorithms to perform feedback consolidation at the branch points. This paper discusses various design options and implementation alternatives for the consolidation algorithms, and proposes a number of new algorithms. The performance of the proposed algorithms and the previous algorithms is compared under a variety of conditions. Results indicate that the algorithms we propose eliminate the consolidation noise (caused if the feedback is returned before all branches respond), while exhibiting a fast transient response.

## 1 Introduction

ABR flow control requires the sources to send at the rate specified by the network in feedback (resource management) cells. For point-to-multipoint connections, feedback consolidation at the branch points becomes necessary. The operation of feedback consolidation can be explained by figure 1. The consolidation operation avoids the feedback implosion problem, where the number of backward resource management (BRM) cells received by the source is proportional to the number of leaves in the multicast tree. In addition, the allowed rate of the source should not fluctuate due to the varying feedback received from different leaves.

In *point-to-point* ABR connections, the source transmits at the minimum rate that can be supported by all the switches on the path from the source to the destination [5]. The natural extension of this strategy for point-to-multipoint connections is controlling the source to the minimum rate that can be supported by the switches on the paths from the source to *all of the leaves* in the multicast tree. The minimum rate is the technique most compatible with typical data requirements: no data should be lost, and the network can take whatever time needed for data delivery.

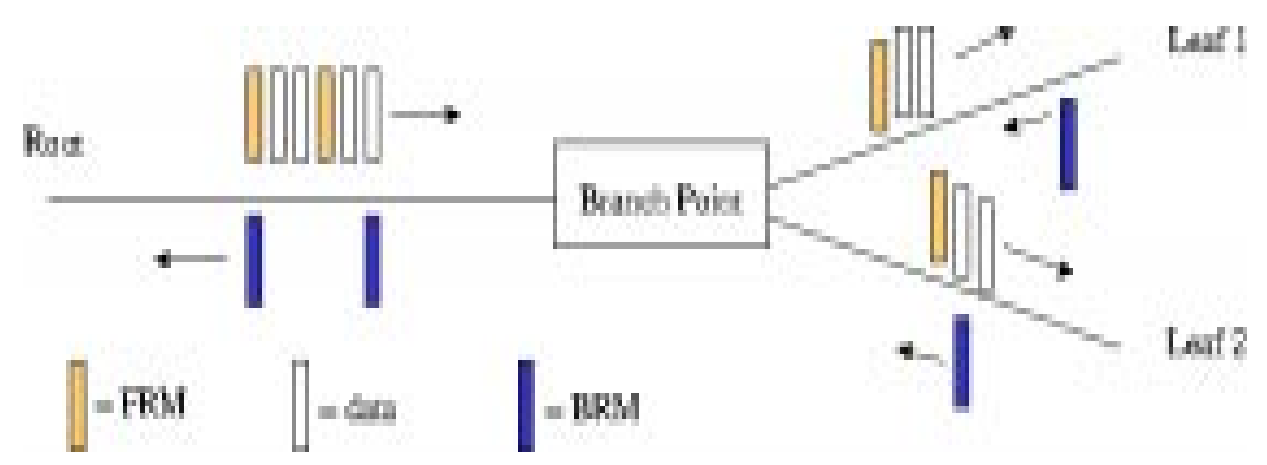

Figure 1: Point-to-multipoint connections

A number of consolidation algorithms have been proposed in [1, 7, 8, 9]. Several design and implementation considerations come into play when developing a consolidation algorithm. The oscillations and transient response of the algorithm are important. The algorithm must also be scalable to very large multicast trees. The implementation complexity, feedback delay, and the overhead of the backward RM cells should not increase with the increase of the number of levels or leaves of the multicast tree.

In this paper, we propose a set of consolidation algorithms that aim at providing a fast transient response, while eliminating consolidation noise. We examine the performance of the proposed algorithms, and compare it to the previous ones in complexity, transient response, consolidation noise, and scalability. The remainder of the paper is organized as follows. The next two sections provide an overview of the ABR flow control mechanism, and a summary of the previous work on point-to-multipoint ABR flow control. A discussion of the various design and implementation issues involved is then presented, followed by a description of the specific underlying switch scheme employed. An explanation and pseudocode of the previously proposed consolidation algorithms, as well as the new ones we propose, is presented next. All the algorithms are then simulated and analyzed under a variety of configurations. The paper concludes with a discussion of the tradeoffs among the algorithms.

*This research was sponsored in part by Rome Laboratory/C3BC Contract # F30602-96-C-0156.

## 2 ABR Flow Control

The available bit rate (ABR) service for data traffic in ATM networks periodically indicates to sources the rate at which they should be transmitting. The switches monitor their load and compute the available bandwidth, dividing it fairly among active flows. The feedback from the switches to the sources is sent in resource management (RM) cells which are generated by the sources and turned around by the destinations.

The RM cells contain the source current cell rate (CCR), in addition to fields that can be used by the switches to provide feedback to the sources. These fields are: explicit rate (ER), the congestion indication (CI) flag, and no increase (NI) flag. The ER field indicates the rate that the network can support at this particular instant. Initially, the ER field is set to a value no greater than the peak cell rate (PCR), and the CI and NI flags are clear. Each switch on the path *reduces* the ER field to the maximum rate it can support, and sets CI or NI if necessary [5].

A component $c_j$ is said to be *downstream* of another component $c_i$ in a certain connection if $c_j$ is on the path from $c_i$ to the destination. In this case, $c_i$ is said to be *upstream* of $c_j$. RM cells flowing from the source to the destination are called forward RM cells (FRMs) while those returning from the destination to the source are called backward RM cells (BRMs). When a source receives a BRM, it computes its allowed cell rate (ACR) using its current ACR, the CI and NI flags, and the ER field of the RM cell.

## 3 Related Work

A simple point-to-multipoint ABR algorithm was proposed in [8]. In this algorithm, a register MER (minimum explicit rate), maintains the minimum feedback indicated by the BRM cells received from the branches. Whenever an FRM cell is received, it is multicast to all branches, and a BRM is returned with the MER value as the explicit rate. MER is then reset.

This algorithm suffers from the "consolidation noise" problem when a BRM generated by a branch point does not consolidate feedback from all tree branches [6]. In fact, if a BRM generated by the branch point does not accumulate feedback from *any* branch, the feedback can be given as the peak cell rate (if that branch point itself is not overloaded). In [9, 7] some solutions to this problem are proposed.

To reduce the complexity of the scheme, [7] also proposes to forward one of the BRM cells returned by the leaves, instead of turning around the FRM cells of the source. Another alternative would be to pass back the BRM cell only when BRM cells from *all* branches have been received after the last feedback. This idea is also used in [1], but the BRM cell that is allowed to pass back to the source is the last BRM cell to be received with a certain sequence number.

## 4 Design Issues

As previously mentioned, there are several ways to implement the consolidation algorithm at branch points. Each method offers a tradeoff in complexity, scalability, overhead, transient response, and consolidation noise. The tradeoffs can be summarized as follows:

[**A**] Which component generates the BRM cells (i.e., turns around the FRM cells)? Should the branch point, or should the destination, perform this operation?

[**B**] Should the branch point wait for feedback from all the branches before passing the BRM cell upstream? Although this eliminates the consolidation noise, it incurs additional complexity, and increases the transient response of the scheme, especially after idle or low rate periods.

[**C**] How can the ratio of FRM cells generated by the source to BRM cells returned to the source be controlled?

[**D**] How can the ratio of BRM cells *in the network* to the source-generated FRM cells be controlled?

[**E**] How does the branch point operate when the it is also a switch and queuing point? The coupling of the switch and branch point functions must be considered. When should the actual rate computation algorithm be invoked?

[**F**] How can the scheme be scalable? Some algorithms wait for an FRM cell to be received to send feedback. Will the feedback delay grow with the number of branches?

[**G**] How is accounting performed at the branch point? Consolidation algorithms use registers to store values such as the minimum rate given by branches in the current iteration, and flags to indicate whether an RM cell has been received since the last one was sent. Some values, such as the minimum explicit rate, should not be stored per output port.

[**H**] How are non-responsive branches handled? If the consolidation scheme waits for feedback from *all* the branches before sending a BRM to the source, an algorithm must be developed to determine when a branch becomes non-responsive and handle this case. Such mechanisms will be the subject of a future study.

## 5 The ERICA Algorithm

The ERICA algorithm is used in our simulations to calculate the explicit rate (ER) feedback in RM cells based on the load at each port. In this section, we only present the basic features of the algorithm. For a more complete explanation of the algorithm, refer to [2][1]. The point-to-multipoint algorithms are presented in the next two sections.

ERICA aims at fair and efficient allocation of the available bandwidth to all contending sources. The ERICA scheme periodically monitors the load on each link and determines a load factor, $z$, the available capacity, and the number of currently active connections (VCs). The load factor, $z$, is an indicator of the congestion level of the link. The optimal op-

[1] All our papers and ATM Forum contributions are available through http://www.cis.ohio-state.edu/~jain/

erating point is at an overload value of one. The load factor is calculated as the ratio of the measured input rate at the port to the target capacity of the output link:

$z$←ABR Input Rate/ABR Capacity

where

ABR Capacity←Target Utilization × Link Bandwidth − VBR Usage − CBR Usage.

Target utilization is a parameter set to a fraction (close to, but less than 100%).

The switch calculates the quantity:

VCShare←$CCR/z$

If all VCs change their rate to their $VCShare$ values then, in the next cycle, the switch would experience unit overload ($z = 1$).

The fair share of each VC, $FairShare$, is also computed as follows:

FairShare←ABR Capacity/Number of Active VCs

A combination of the two quantities $FairShare$ and $VCShare$ is used to rapidly reach optimal operation as follows:

ER Calculated← Max (FairShare, VCShare, Maximum ER in previous interval)

Several enhancements to this algorithm avoid transient overloads, and take the queuing delay into consideration when assessing the available capacity. Averaging the measured quantities further improves the performance. These enhancements are described in [2].

# 6 Consolidation Algorithms

This section describes some previously proposed consolidation algorithms, while the next section proposes a number of new algorithms. In the algorithms presented, ERICA (as explained in the previous section) is employed immediately before sending a BRM on the link. This ensures that the most recent feedback information is sent. The algorithms at the branch point operate as explained in the following subsections.

## 6.1 Algorithm 1

This algorithm is a modified version of the algorithm in [8]. The main idea of the algorithm is that BRM cells are returned from the branch point when FRM cells are received, and contain the minimum of the values indicated by the BRM cells received from the branches after the last BRM cell was sent. FRM cells are duplicated and multicast to all branches at the branch point.

A register, MER, and two flags, MCI and MNI, are maintained for each multipoint VC. The variables store the minimum of the explicit rate (ER), congestion indication (CI) and no increase (NI) indicated in the BRM cells which were received after the last BRM cell was sent. MER is initialized to the peak cell rate, while CI and NI are initialized to zero. Three temporary variables: MXER, MXCI, and MXNI are also used when an FRM cell is received (their values do not persist across invocations of the algorithm). They store the ER, CI and NI from the FRM cell. The algorithm operates as follows.

**Upon the receipt of an FRM cell:**

1. Multicast FRM cell to all participating branches
2. Let MXER = ER from FRM cell, MXCI = CI from FRM cell, MXNI = NI from FRM cell
3. Return a BRM with ER = MER, CI = MCI, NI = MNI to the source
4. Let MER = MXER, MCI = MXCI, MNI = MXNI

**Upon the receipt of a BRM cell:**

1. Let MER = min (MER, ER from BRM cell), MCI = MCI OR CI from BRM cell, MNI = MNI OR NI from BRM cell
2. Discard the BRM cell

**When a BRM is about to be scheduled:**

Let ER = min (ER, ER calculated by ERICA for all branches)

## 6.2 Algorithm 2

This algorithm is a modified version of the second algorithm in [7]. The only change from Algorithm 1 (as described above) is ensuring that at least one BRM cell has been received from a branch before turning around an FRM. For this purpose, a boolean flag, AtLeastOneBRM (initially zero), is set to true when a BRM cell is received from a branch, and reset when a BRM is sent by the branch point. As before, MER, MCI, MNI, and here, AtLeastOneBRM, are stored for each multipoint VC, and MXER, MXCI, MXNI are temporary variables.

**Upon the receipt of an FRM cell:**

1. Multicast FRM cell to all participating branches
2. IF AtLeastOneBRM THEN
   A. Let MXER = ER from FRM cell, MXCI = CI from FRM cell, MXNI = NI from FRM cell
   B. Return a BRM with ER = MER, CI = MCI, NI = MNI to the source
   C. Let MER = MXER, MCI = MXCI, MNI = MXNI
   D. Let AtLeastOneBRM = 0

**Upon the receipt of a BRM cell:**

1. Let AtLeastOneBRM = 1
2. Let MER = min (MER, ER from BRM cell), MCI = MCI OR CI from BRM cell, MNI = MNI OR NI from BRM cell
3. Discard the BRM cell

**When a BRM is about to be scheduled:**

Let ER = min (ER, ER calculated by ERICA for all branches)

## 6.3 Algorithm 3

This is a modified version of the third algorithm in [7]. The main idea here is that the branch point does not turn around the FRMs, but the BRM that is received from a branch immediately after an FRM has been received by the branch point is passed back to the source, carrying the minimum values. A boolean flag, AtLeastOneFRM, indicates that an FRM cell has been received by the branch point after the last BRM cell was passed to the source. Again, MER, MCI, MNI, and AtLeastOneFRM are stored per multipoint VC.

**Upon the receipt of an FRM cell:**
1. Multicast FRM cell to all participating branches
2. Let AtLeastOneFRM = 1

**Upon the receipt of a BRM cell:**
1. Let MER = min (MER, ER from BRM cell), MCI = MCI OR CI from BRM cell, MNI = MNI OR NI from BRM cell
2. IF AtLeastOneFRM THEN
   A. Pass the BRM with ER = MER, CI = MCI, NI = MNI to the source
   B. Let MER = PCR, MCI = 0, MNI = 0
   C. Let AtLeastOneFRM = 0
   ELSE Discard the BRM cell

**When a BRM is about to be scheduled:**
Let ER = min (ER, ER calculated by ERICA for all branches)

### 6.4 Algorithm 4

A variation of this algorithm was presented in [7] as algorithm 4, and another variation using sequence numbers in RM cells was proposed in [1]. The main idea here is that a BRM is passed to the source only when BRM cells have been received from all branches. To count the number of branches from which BRM cells were received at the branch point (after the last BRM cell was passed by the branch point), a counter, NumberOfBRMsReceived is incremented the first time a BRM cell is received from each branch (NumberOfBRMsReceived is initialized to zero). As before, the MER, MCI, MNI, and NumberOfBRMsReceived registers are maintained per multipoint VC. The value of the NumberOfBRMsReceived counter is compared to the value of another counter, NumberOfBranches, every time a BRM cell is received by the branch point. If the value of NumberOfBRMsReceived is equal to NumberOfBranches, the BRM cell is passed back to the source, carrying the values of the MER, MCI and MNI registers. [NumberOfBranches stores the number of branches of the point-to-multipoint VC at this branch point. It is also stored for each VC, and initialized during connection setup. In addition, if leaf initiated join is allowed (as in UNI 4.0), NumberOfBranches must be updated every time a branch is added to a branch point.]

A flag, BRMReceived, is needed for each branch to indicate whether a BRM cell has been received from this particular branch, after the last BRM cell was passed. The flag is stored for each output port and not for each VC, since it is needed for each branch. Note that a timeout mechanism must be implemented to ensure that BRM cell flow is not stopped in the case of non-responsive branches.

**Upon the receipt of an FRM cell:**
Multicast FRM cell to all participating branches

**Upon the receipt of a BRM cell from branch $i$:**
1. IF NOT $\text{BRMReceived}_i$ THEN
   A. Let $\text{BRMReceived}_i = 1$
   B. Let NumberOfBRMsReceived = NumberOfBRMsReceived + 1
2. Let MER = min (MER, ER from BRM cell), MCI = MCI OR CI from BRM cell, MNI = MNI OR NI from BRM cell
3. IF NumberOfBRMsReceived is equal to NumberOfBranches THEN
   A. Pass the BRM with ER = MER, CI = MCI, NI = MNI to the source
   B. Let MER = PCR, MCI = 0, MNI = 0
   C. Let NumberOfBRMsReceived = 0
   D. Let BRMReceived = 0 FOR all branches
   ELSE Discard the BRM cell

**When a BRM is about to be scheduled:**
Let ER = min (ER, ER calculated by ERICA for all branches)

## 7 New Algorithms

The main problem with algorithm 4 described in the previous section is its slow transient response. Even when excessive overload has been detected, the algorithm has to wait for feedback from (possibly distant) leaves before indicating the overload information to the source. By that time, the source might have transmitted a large number of cells (which would be dropped due to buffer overflow), leading to performance degradation. This situation is especially problematic when the source has been idle for some time, and then suddenly sends a burst, so there are no RM cells initially in the network.

The main idea behind the algorithms presented next is that the slow transient response problem should be avoided when an overload situation has been detected. In this case, there is no need to wait for feedback from all the branches, and the overload should be immediately indicated to the source. In cases of underload indication from a branch, it is better to wait for feedback from all branches, since other branches may be overloaded. This is somewhat similar to the idea behind the backward explicit congestion notification (BECN) cells sent by the switches.

Overload is detected when the feedback to be indicated is *much less* than the last feedback returned by the branch point (the "much less" condition can be tested using a multiplicative factor). An alternative method would be to compare the feedback to be indicated to the *current cell rate (CCR)* or ACR of the VC. Although this may be better because it accounts for upstream bottlenecks, and prevents the transmission of unnecessary BRM cells in such cases, the CCR information may be stale due to the delay from the source to the branch point (it may also be much larger when the source becomes idle or becomes a low rate source after the last FRM was sent), and a large number of BRMs may be sent in such cases. The last feedback indicated by the branch point is a more current value. The minimum of the CCR and last feedback given can be used in the comparison, but this involves some additional complexity, and may slow down the overload response when the CCR happens to have been small, but is currently large.

Note that when a BRM cell is returned due to overload detection before feedback has been received from all branches, the counters and the register values are not reset.

## 7.1 Fast Overload Indication (Algorithm 5)

In this algorithm, the LastER register maintains the last explicit rate value returned by the branch point (LastER is initialized to the initial cell rate (ICR) of the connection). LastER is stored per multipoint VC.

Two temporary variables: SendBRM and Reset are used. SendBRM is set only if a BRM cell is to be passed to the source by the branch point. Reset is false only if a BRM cell is being used to indicate overload conditions, and hence the register values should not be reset. FRMminusBRM is only used for accounting purposes, and will not exist in a real implementation.

**Upon the receipt of an FRM cell:**

1. Multicast FRM cell to all participating branches
2. Let FRMminusBRM = FRMminusBRM + 1

**Upon the receipt of a BRM cell from branch $i$:**

1. Let SendBRM = 0
2. Let Reset = 1
3. IF NOT BRMReceived$_i$ THEN
    A. Let BRMReceived$_i$ = 1
    B. Let NumberOfBRMsReceived = NumberOfBRMsReceived + 1
4. Let MER = min (MER, ER from BRM cell), MCI = MCI OR CI from BRM cell, MNI = MNI OR NI from BRM cell
5. IF MER << LastER THEN (* overload is detected *)
    A. IF NumberOfBRMsReceived < NumberOfBranches THEN
        1. Let Reset = 0
    B. Let SendBRM = 1
    ELSE IF NumberOfBRMsReceived is equal to NumberOfBranches THEN
    A. Let SendBRM = 1
6. IF SendBRM THEN
    A. Pass the BRM with ER = MER, CI = MCI, NI = MNI to the source
    B. IF Reset THEN
        1. Let MER = PCR, MCI = 0, MNI = 0
        2. Let NumberOfBRMsReceived = 0
        3. Let BRMReceived = 0 FOR all branches
    C. Let FRMminusBRM = FRMminusBRM − 1
    ELSE Discard the BRM cell

**When a BRM is about to be scheduled:**

1. Let ER = min (ER, ER calculated by ERICA for all branches)
2. Let LastER = ER

## 7.2 RM Ratio Control (Algorithm 6)

The previous algorithm may increase the BRM cell overhead, since the ratio of source-generated FRM cells to BRM cells received by the source can be more than one. To avoid this problem, we introduce the register SkipIncrease which is maintained for each multipoint VC (and initialized to zero). SkipIncrease is incremented whenever a BRM cell is sent before feedback from all the branches has been received. When feedback from all leaves indicates underload, and the value of the SkipIncrease register is more than zero, this particular feedback can be ignored and SkipIncrease is decremented. Note that the value of the SkipIncrease counter will not increase to large values, since the congestion avoidance algorithm (such as ERICA) arrives at the optimal allocation within few iterations, and the rate allocations *cannot continue decreasing indefinitely*. Our analysis and simulations have shown that the counter never exceeds small values and quickly stabilizes at zero. A maximum value can also be enforced by the algorithm.

**Upon the receipt of an FRM cell:**

1. Multicast FRM cell to all participating branches
2. Let FRMminusBRM = FRMminusBRM + 1

**Upon the receipt of a BRM cell from branch $i$:**

1. Let SendBRM = 0
2. Let Reset = 1
3. IF NOT BRMReceived$_i$ THEN
    A. Let BRMReceived$_i$ = 1
    B. Let NumberOfBRMsReceived = NumberOfBRMsReceived + 1
4. Let MER = min (MER, ER from BRM cell), MCI = MCI OR CI from BRM cell, MNI = MNI OR NI from BRM cell
5. IF MER $\geq$ LastER AND SkipIncrease > 0 AND NumberOfBRMsReceived is equal to NumberOfBranches THEN
    A. Let SkipIncrease = SkipIncrease − 1
    B. Let NumberOfBRMsReceived = 0
    C. Let BRMReceived = 0 FOR all branches
    ELSE IF MER << LastER THEN
    A. IF NumberOfBRMsReceived < NumberOfBranches THEN
        1. Let SkipIncrease = SkipIncrease + 1
        2. Let Reset = 0
    B. Let SendBRM = 1
    ELSE IF NumberOfBRMsReceived is equal to NumberOfBranches THEN
    A. Let SendBRM = 1
6. IF SendBRM THEN
    A. Pass the BRM with ER = MER, CI = MCI, NI = MNI to the source
    B. IF Reset THEN
        1. Let MER = PCR, MCI = 0, MNI = 0
        2. Let NumberOfBRMsReceived = 0
        3. Let BRMReceived = 0 FOR all branches
    C. Let FRMminusBRM = FRMminusBRM − 1
    ELSE Discard the BRM cell

**When a BRM is about to be scheduled:**

1. Let ER = min (ER, ER calculated by ERICA for all branches)
2. Let LastER = ER

### 7.3 Immediate Rate Calculation (Algorithm 7)

The previously discussed algorithms can offer very fast congestion relief when an overload is detected in a branch of the multicast tree. However, they do not account for the potential overload situation at the branch point itself, since if the branch point is a switch, the ERICA algorithm is only performed when the BRM cell is about to be scheduled on the link. In cases when the branch point is itself a switch and a queuing point, the immediate rate calculation option invokes ERICA whenever a BRM is received, and not just when a BRM is being sent. Hence overload at the branch point can be detected and indicated according to the fast overload indication option as previously described. This option, however, may involve some additional complexity.

The algorithm presented next is the same as Algorithm 6 in the previous subsection, except for the addition of the ERICA invocation (italicized below).

**Upon the receipt of an FRM cell:**

1. Multicast FRM cell to all participating branches
2. Let FRMminusBRM = FRMminusBRM + 1

**Upon the receipt of a BRM cell from branch $i$:**

1. Let SendBRM = 0
2. Let Reset = 1
3. IF NOT BRMReceived$_i$ THEN
   A. Let BRMReceived$_i$ = 1
   B. Let NumberOfBRMsReceived = NumberOfBRMsReceived + 1
4. Let MER = min (MER, ER from BRM cell), MCI = MCI OR CI from BRM cell, MNI = MNI OR NI from BRM cell
5. *Let MER = min (MER, minimum ER calculated by ERICA for all branches)*
6. IF MER $\geq$ LastER AND SkipIncrease > 0 AND NumberOfBRMsReceived is equal to NumberOfBranches THEN
   A. Let SkipIncrease = SkipIncrease − 1
   B. Let NumberOfBRMsReceived = 0
   C. Let BRMReceived = 0 FOR all branches
   ELSE IF MER << LastER THEN
   A. IF NumberOfBRMsReceived < NumberOfBranches THEN
      1. Let SkipIncrease = SkipIncrease + 1
      2. Let Reset = 0
   B. Let SendBRM = 1
   ELSE IF NumberOfBRMsReceived is equal to NumberOfBranches THEN
   A. Let SendBRM = 1
7. IF SendBRM THEN
   A. Pass the BRM with ER = MER, CI = MCI, NI = MNI to the source
   B. IF Reset THEN
      1. Let MER = PCR, MCI = 0, MNI = 0
      2. Let NumberOfBRMsReceived = 0
      3. Let BRMReceived = 0 FOR all branches
   C. Let FRMminusBRM = FRMminusBRM − 1
   ELSE Discard the BRM cell

**When a BRM is about to be scheduled:**

1. Let ER = min (ER, ER calculated by ERICA for all branches)
2. Let LastER = ER

## 8 Performance Analysis

This section provides a performance comparison among all the consolidation algorithms, in a variety of configurations with bursty and non-bursty traffic, with and without variable bit rate (VBR) background, and with various link lengths, bottleneck locations, and number of leaves. A large number of other configurations was also tested (see [3, 4] for some of the configurations), but only a sample of the results is shown here. In particular, configurations with a large number of leaves at varying distances in the multicast tree were simulated, and the results were consistent.

### 8.1 Parameter Settings

Throughout our experiments, the following parameter values are used:

[**1**] All links have a bandwidth of 155.52 Mbps (149.76 Mbps when SONET overhead is accounted for).

[**2**] All point-to-multipoint traffic flows from the root to the leaves of the tree. No traffic flows from the leaves to the root, except for RM cells. The same applies for point-to-point connections.

[**3**] All sources are deterministic, i.e., their start/stop times and their transmission rates are known. The bursty traffic sources send data in bursts, where each burst starts after a small request has been received from the client. VBR sources are on for 20 ms and off for 20 ms.

[**4**] The source parameter rate increase factor (RIF) is set to one, to allow immediate use of the full explicit rate indicated in the returning RM cells at the source. Initial cell rate (ICR) is also set to a high value (almost peak cell rate). These factors are set to such high values to simulate a *worst case* load situation.

[**5**] The source parameter transient buffer exposure (TBE) is set to large values to prevent rate decreases due to the triggering of the source open-loop congestion control mechanism. This was done to isolate the rate reductions due to the switch congestion control from the rate reductions due to TBE.

[**6**] The switch target utilization parameter was set at 90%. The switch measurement interval was set to the minimum of the time to receive 100 cells and 1 ms.

### 8.2 Simulation Results

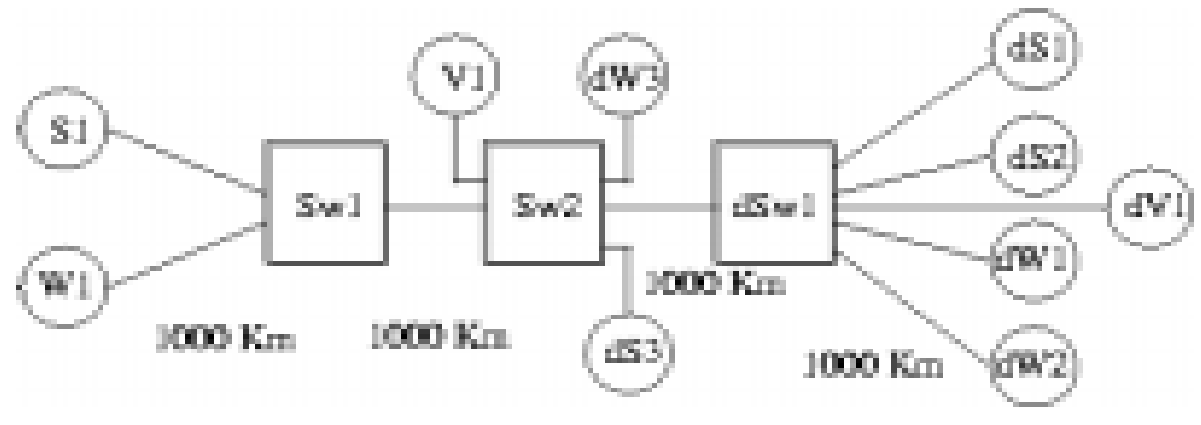

Figure 2: WAN parking lot configuration with bursty, infinite and VBR connections

This section discusses the performance of the seven consolidation algorithms by comparing them in a set of configurations. First, the performance of the seven different algorithms was tested in a situation where there is both variable capacity and variable demand. These situations offer the toughest challenge for rate allocation algorithms [4]. The first configuration simulated is shown in figure 2. The source indicated by W is a bursty source, I is a persistent (infinite) source, while V is a VBR source. Note that the high initial cell rate (ICR) and rate increase factor (RIF) [5] values are the reason for the unusually large queues seen for all algorithms.

The source ACR graphs (not shown here) for algorithms 1, 2, and 3 indicate many fluctuations and inaccurate (around 140 Mbps) feedback given in the initial 150 ms. This leads to large queues (>5000 cells with every VBR burst). Algorithm 4 gives more accurate feedback, but the feedback is given after around 50 ms, which results in initially large queues (since ICR is large). Algorithms 5 and 6 produce identical results to algorithm 4, since the bottleneck link is attached to the branch point. Algorithm 7, on the other hand, exhibits a very fast transient response, and gives relatively accurate feedback to both sources. The initial queues caused by high ICR, as well as the queues with every VBR burst are much smaller). Hence, it offers the best performance since it combines the benefits of algorithm 4 with a fast transient response.

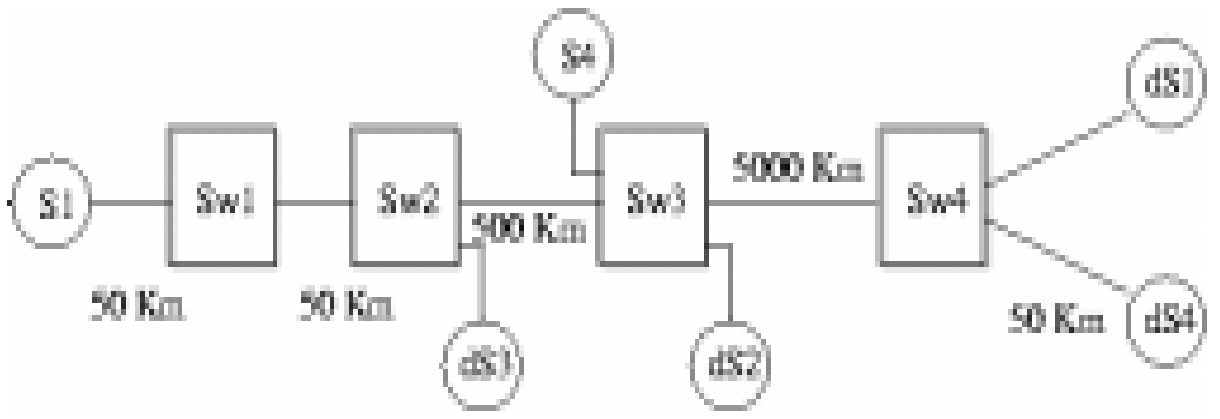

Figure 3: Chain configuration

The chain configuration, illustrated in figure 3 consists of a point-to-multipoint connection where one of the links on the route to the farthest leaf is the bottleneck link. Also the link lengths increase by an order of magnitude in each of the last two hops.

As seen in figures 4 through 10, this configuration is an ideal configuration for illustrating the consolidation noise problem. The problem is severe for algorithms 1, 2 and 3 (especially 3) (see figures 4 through 6), and leads to rate oscillations, instability, unbounded queues and unfairness against source S4 whose rate remains at half of the bandwidth, while the rate of S1 continues to oscillate around a mean of about 103 Mbps. Although using a scheme such as ERICA+ leads to stability and bounded queues in this case, the persistent rate oscillations result in unacceptable performance and unfairness (*the problem can be mitigated by using small RIF values, but this slows down rate increases*). Algorithms 4, 5 and 6 (figures 7 through 9) avoid the noise completely, but suffer from a slow transient response. The rate of the source S1 only drops after around 60 ms, and by that time, large queues have built up at the switches). Algorithm 7 yields optimal performance in this case, since the rate of the source S1 immediately drops to its optimal value, as soon as the overload is detected.

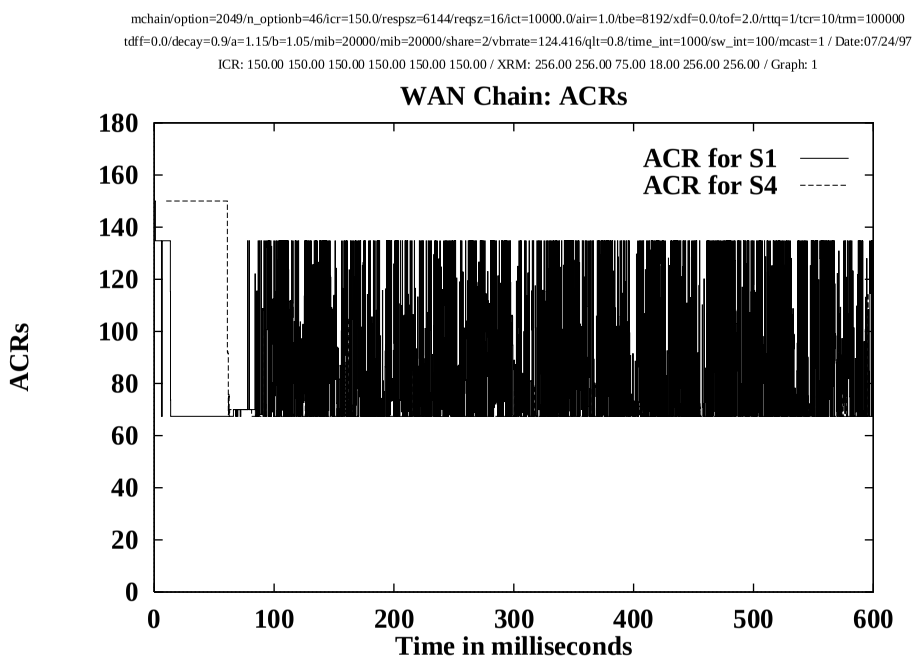

Figure 4: Results for a Chain configuration [Algorithm 1]

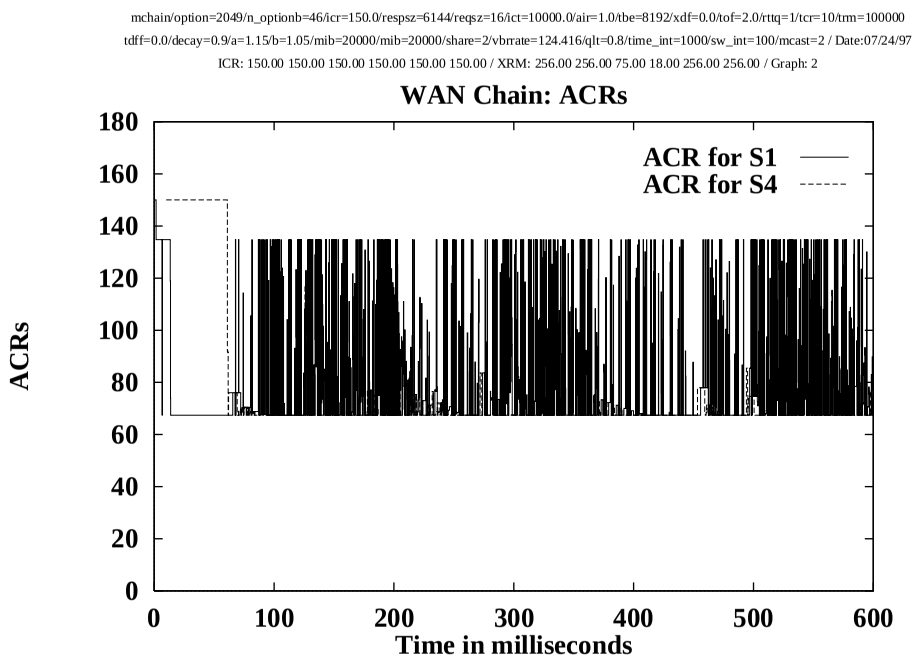

Figure 5: Results for a Chain configuration [Algorithm 2]

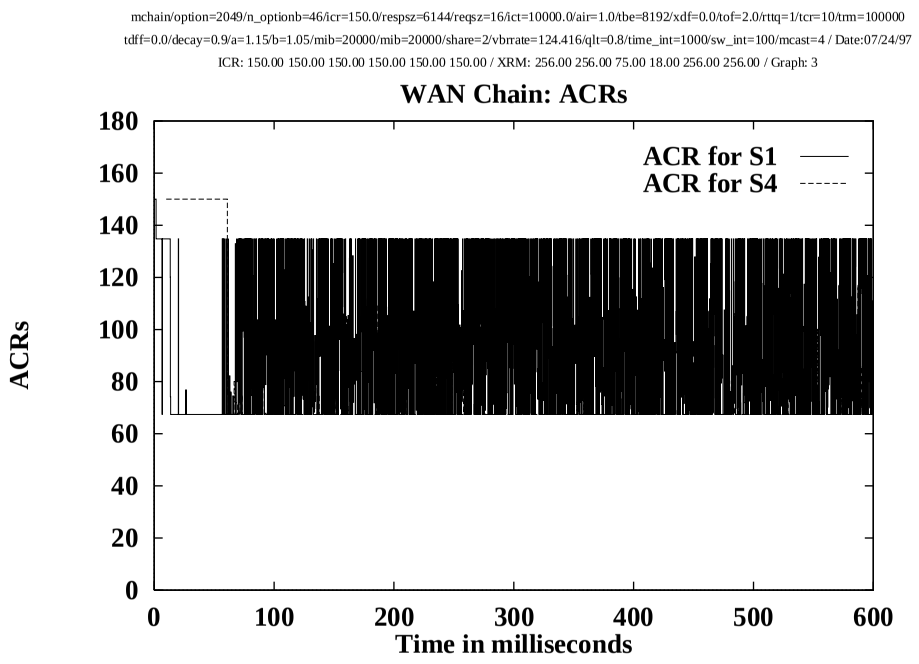

Figure 6: Results for a Chain configuration [Algorithm 3]

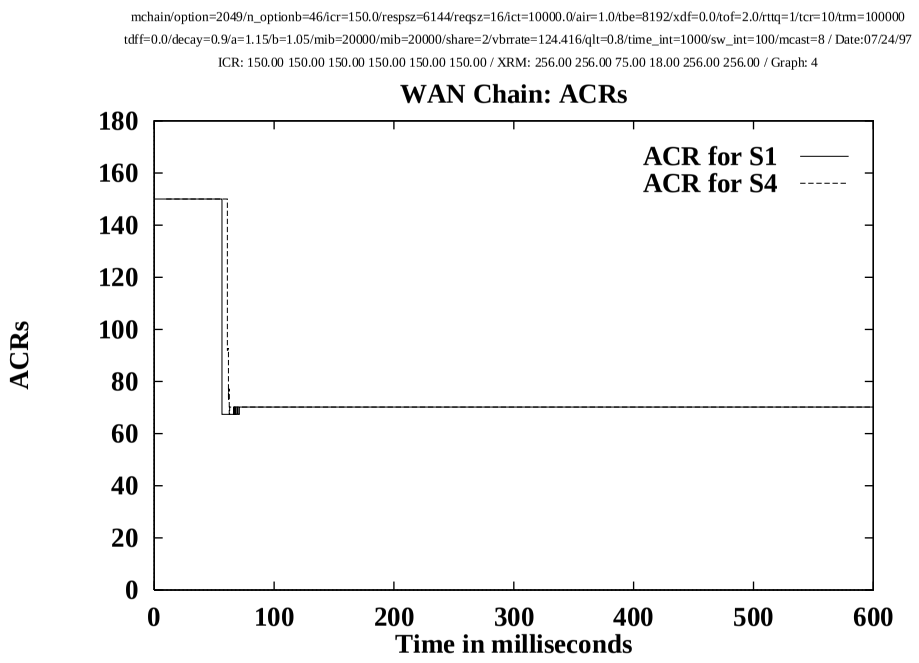

Figure 7: Results for a Chain configuration [Algorithm 4]

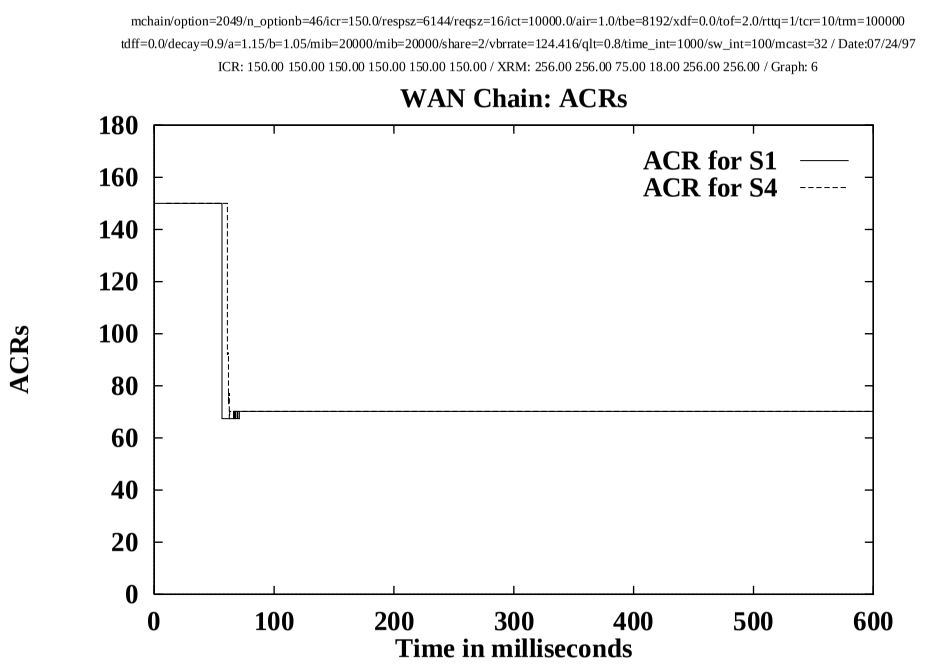

Figure 8: Results for a Chain configuration [Algorithm 5]

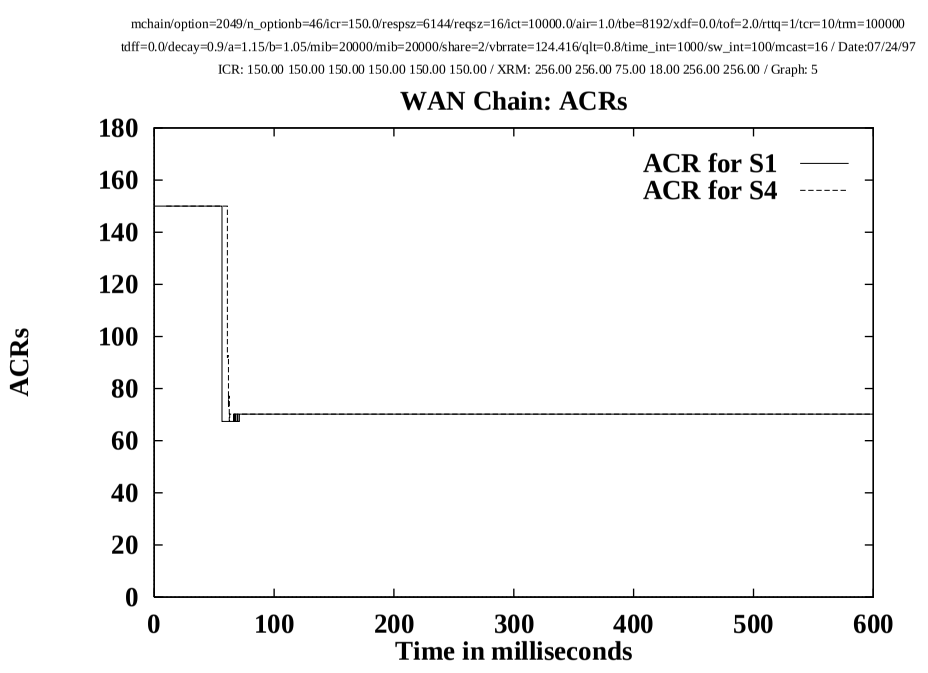

Figure 9: Results for a Chain configuration [Algorithm 6]

Observe that algorithms 5 and 6 also yield near optimal performance (like algorithm 7) if the destination dS3 was further than dS1, as in the configuration in figure 11. Here, the chain configuration is modified such that the bottleneck link is closer to the branch point at switch Sw2 than another leaf, namely dS3.

In this case (figure is not shown here), algorithm 4 wastes a long time waiting for feedback from dS3, while it has already received the bottleneck feedback from Sw3. Algorithms 5, 6, and 7 send the feedback as soon as the overload situation is indicated by the BRM cell coming from switch Sw3, and do not needlessly wait for the BRM from dS3. Hence, the 3 new algorithms perform near optimally since the rate of the source S1 goes to the optimal value after only around 20 ms for algorithms 5 and 6, and less than 10 ms for algorithm 7. The maximum queue length is also much smaller than for algorithm 4 ($>$ 16000 cells): for algorithms 5 and 6, it is around 7000 cells, and for algorithm 7, it is only 3500 cells.

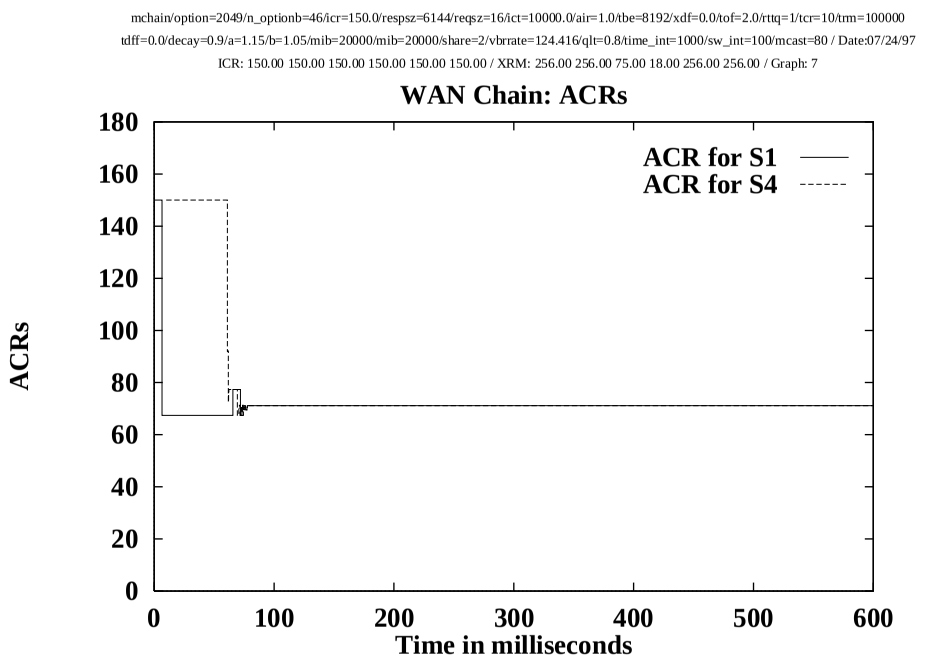

Figure 10: Results for a Chain configuration [Algorithm 7]

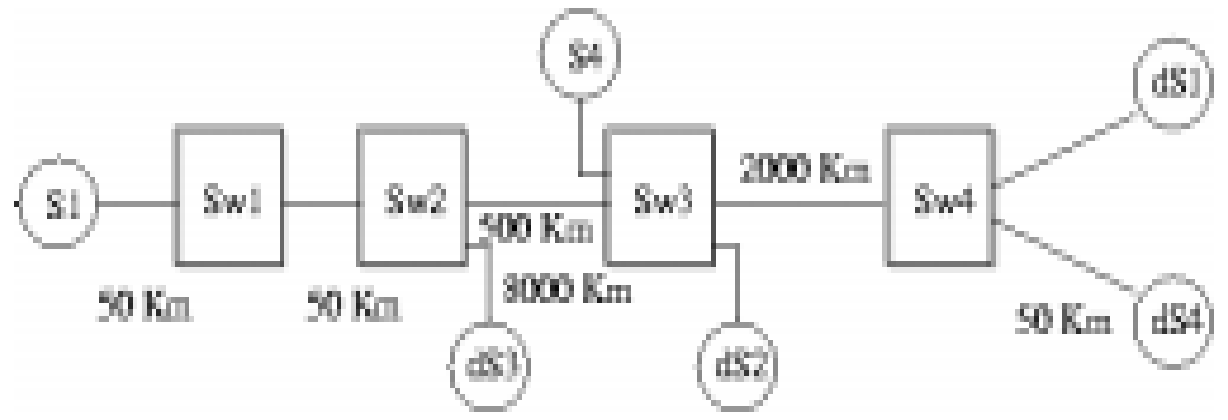

Figure 11: Modified chain configuration

We observed a similar, but more pronounced, behavior when we simulated configurations with a larger number of leaves at varying distances and at varying levels of the multicast tree. The situation was much worse in those cases with algorithms 1, 2, and 3, which had much more severe noise problems. Algorithm 4 had an *extremely* slow transient response, while algorithms 5, 6, and *especially 7* quickly reached the optimal values, and the queues at the switches were small.

# 9 Comparison of the Algorithms

This section summarizes the conclusions from the comparison of the various algorithms. All the algorithms preserve the fairness and efficiency of the point-to-point congestion avoidance algorithm employed. We compare the space and time complexity, transient response, consolidation noise, algorithm overhead and scalability, and discuss the interoperability of various algorithms.

## 9.1 Implementation Complexity

Algorithms 1 and 2 are complex to implement because the branch point has to turn around the RM cells. This is somewhat similar to the Virtual Source/Virtual Destination (VS/VD) concept. Most studies argue that turning around RM cells has a high implementation complexity.

Algorithm 3 is definitely the simplest algorithm to implement, since it does not turn around RM cells, and it keeps minimal per-VC accounting information. Algorithm 4 is more complex since it has to maintain the number of branches and the number of branches from which BRMs have been received, and compare those numbers. In addition, it has to maintain a bit for each output port to denote whether a BRM cell has been received from this branch, and some timeout-related values.

Algorithms 5 and 6 are slightly more complex since they may also store the last ER sent by the branch point. Alternatively, they can use the CCR of the source, which is already stored and used by most congestion avoidance algorithms (it is used in the ERICA algorithm which we have employed in this study). Hence, the additional complexity mainly stems from the comparison of the MER value to the last ER sent or the CCR value, and maintaining the SkipIncrease counter. The additional comparison and integer register do not incur much overhead.

Algorithm 7 is somewhat more complex than algorithms 5 and 6, since it invokes the ERICA algorithm for all the branches whenever a BRM cell is received by the branch point, and not only when a BRM cell is to be sent.

## 9.2 Transient Response

Algorithm 1 exhibits a very fast transient response. Algorithms 2 and 3 also have a reasonable transient response, since, even if there are no RM cells in the network, the feedback is quickly returned on the first BRM arrival.

Algorithm 4 has a slow transient response, since it waits for feedback from all the leaves before sending BRMs. This is especially severe in cases when there are few or no RM cells already in the network, such as during startup periods and for bursty sources. Therefore feedback can be delayed up to a function of the longest round trip times of the leaves. Algorithms 5, 6 and 7 tackle this problem for overload situations. The transient response of the schemes is very fast when an overload is detected downstream (for algorithms 5 and 6), or at this branch and downstream (for algorithm 7). In such cases, the transient response of the scheme is reasonably fast, and potential cell loss and retransmissions are alleviated.

## 9.3 Consolidation Noise

Algorithms 1, 2, and 3 suffer from severe consolidation noise problems. In particular, algorithms 1 and 3 suffer from unacceptable consolidation noise in some cases (recall figures 4 and 6). Algorithm 2 somewhat alleviates these problems, since BRMs are not sent if no feedback has been received from any of the downstream components. However, it still exhibits considerable noise.

Algorithms 4, 5, 6, and 7 eliminate this problem by waiting for feedback from all branches. Although algorithms 5, 6, and 7 do not wait for feedback from all leaves in cases of overload, this *does not* introduce noise, since the RM cells that are sent faster than the usual cells carry overload information, which would have been conveyed by the next minimum value anyway.

## 9.4 Scalability Issues

Algorithms should be scalable in the sense that their overhead and feedback delay should not grow with the growth of the number of branch points or levels of the multicast tree.

### 9.4.1 RM cell overhead

The number of FRM cells generated by the source and the number of BRM cells received by the source should be approximately the same. Algorithm 1 generates a BRM cell at the branch point for every FRM cell it receives, thereby guaranteeing that the BRM to FRM ratio remains one. Algorithms 2 and 3 maintain a BRM to FRM ratio of less than or equal to 1 as follows. Algorithm 2 generates a BRM for an FRM only if a BRM has been received from a leaf since the last BRM was sent by the branch point. Algorithm 3 allows a BRM to pass to the source only if an FRM cell has been received by the branch point after the last BRM cell was forwarded. Therefore both algorithms maintain a ratio that is less than or equal to one (actually, it is strictly less than one for algorithm 2, since the first FRM cell will never be turned).

Algorithm 4 also maintains a ratio of less than or equal to one, since one BRM cell is returned when BRM cells have been received from all branches. Algorithm 5 does not guarantee that the ratio remains at 1, since RM cells carrying overload indication are allowed to quickly return to the source. Algorithms 6 and 7 fix this problem by maintaining a counter that is incremented for every extra RM cell passed, and then decremented (and the BRM cell discarded) in cases of RM cells carrying underload information, if the counter exceeds zero. Hence, over the long run, the ratio is maintained at one. The counter cannot increase indefinitely, since the rates cannot decrease indefinitely. In all cases we have examined, the counter value was always small, because ERICA quickly converged.

In addition to the BRM to FRM ratio at the source, the number of BRM cells generated *in the network* per source-generated FRM cell should be controlled. In algorithms 1 and 2, the switch turns around the FRM cells and produces BRM cells, but the same FRM cells are multicast to other branch points and to the leaves, and these also turn around the FRMs. Hence, the number of BRMs in the network can grow with the increase of the number of branch points. This is undesirable. Algorithms 3, 4, 5, 6, and 7 solve this problem, since switches do not turn around the FRM cells.

### 9.4.2 Sensitivity to the maximum number of branch points on a path (levels of the tree)

Algorithm 1 waits for an FRM cell to arrive before it can send the feedback information it has consolidated from the BRM cells. This has to be done at every branch point, leading to a delay that increases with the number of levels of the multicast tree. Algorithm 2 suffers from the same drawback, since the algorithm also sends a BRM cell at the branch point when an FRM cell is received.

Algorithm 3 is less sensitive to the number of levels of the multicast tree. The BRM cell is passed to the source only if an FRM cell has been received since the last BRM cell was sent by the branch point. However, it is passed without additional delay.

Algorithms 4, 5, 6, and 7 are slightly sensitive to the multicast tree levels since BRM cells from all branches are consolidated at every branch point. However, the delay (the time between the transmission of the FRM cell at the source until the source receives the corresponding BRM cell) is mainly dependent on the round trip times from the source to the leaves at that particular time. The round trip times to the leaves can vary with time, dependent on the queuing delay of the switches on the path of the multicast tree. More than one leaf can have an effect on that delay since BRM cells arrive asynchronously at the branch points.

Table 1: Comparison of consolidation algorithms

| Algorithm | 1 | 2 | 3 | 4 | 5 | 6 | 7 |
|---|---|---|---|---|---|---|---|
| Complexity | **High** | **High** | Low | Medium | >Medium | >Medium | >>Medium |
| Transient Response | Fast | Medium | Medium | **Slow** | Fast for overload | Fast for overload | Very fast for overload |
| Noise | High | Medium | **High** | Low | Low | Low | Low |
| BRM:FRM at Root | 1 | <1 | ≤1 | ≤1 | **may be>1** | $lim$ =1 | $lim$ =1 |
| Branch point sensitivity | High | High | Low | Medium | >Medium | Medium | Medium |

### 9.5 Interoperability Issues

The various consolidation algorithms should be able to interoperate with each other if no one algorithm is standardized. It seems that all the algorithms can interoperate smoothly with each other, but the performance of a network with different algorithms at the different branch points, and point-to-multipoint VCs that branch at several branch points with different algorithms, will need further study if a consolidation algorithm is not standardized. This will be one of the areas of our future research work.

## 10 Conclusions

Table 1 shows a summary of the results of the comparison between the consolidation algorithms. In terms of complexity, algorithm 3 is clearly the simplest. Algorithms 1 and 2 are simple, except that the RM turn around operation is expensive. Algorithm 4 introduces additional complexity to algorithm 3, since it maintains per-branch variables and performs comparisons. Algorithm 5 introduces a little more complexity to 4; algorithm 6 introduces a little more to 5; and algorithm 7 introduces some more to 6, but most of the increments are of little complexity.

The transient response of algorithm 1 is fast, but can be erroneous. Algorithms 2 and 3 offer medium response, while algorithm 4 is clearly slow. Algorithms 5, 6, and especially 7, have a fast response when overload is detected. Consolidation noise is a problem with algorithms 1, 2, and 3, especially 1 and 3. The other algorithms overcome this problem.

As for RM cell overhead, the ratio of BRM cells received by the source to FRM cells sent by the source is maintained at unity by algorithm 1. It is less than one for algorithm 2 (at least the first FRM is not returned), and is less than or equal to one for algorithms 3 and 4. Algorithm 5 introduces additional BRM cells in case of overload, while algorithms 6 and 7 ensure the ratio is one over the long run ($lim$ in the table means the limit as time goes to infinity).

Finally, the sensitivity of algorithms 1 and 2 to the number of branch points and the levels of the multicast tree is high due to the additional delay waiting for an FRM cell at each branch point, and the additional BRM cells that are turned around at each level. Algorithms 3 to 7 (especially algorithm 3) are somewhat less sensitive to this.

The comparison clearly indicates that algorithms 1 and 2 suffer from problems. Algorithm 3 is good, except for the consolidation noise problems which lead to unacceptable performance as seen in figure 6. Algorithm 4 provides reasonable performance, but has a slow transient response, which is overcome by the algorithms we proposed (5, 6 and 7). Algorithm 4 and the new algorithms are slightly more complex than algorithm 3, but this can be well worth the performance benefits gained, especially with algorithm 7. *Algorithm 7* avoids congestion, while eliminating the consolidation noise problem.